\documentstyle[twocolumn,aps,epsfig]{revtex}
\begin{document}
\twocolumn[\hsize\textwidth\columnwidth\hsize\csname
@twocolumnfalse\endcsname

\title{Recoil Order Chiral Corrections to Baryon Octet Axial Currents
and Large $N_c$ QCD}
\author{ Shi-Lin Zhu$^{1}$, G. Sacco$^{1,2}$, and
M. J. Ramsey-Musolf$^{1,2}$\\
$^1$ Kellogg Radiation Laboratory, California Institute of Technology,
Pasadena,CA 91125\\
$^2$ Department of Physics, University of Connecticut,
Storrs, CT 06269 \\
}

\maketitle
\begin{abstract}

We compute the chiral corrections to octet baryon axial currents through
${\cal O}(p^3)$ in heavy baryon chiral perturbation theory, 
including both octet and decuplet baryon intermediate states.
We include the latter in a consistent way by using the small scale expansion.
We find that, in contrast to the situation at ${\cal O}(p^2)$, there exist
no cancellations between octet and decuplet contributions at ${\cal O}(p^3)$.
Consequently, the ${\cal O}(p^3)$ corrections spoil the expected scaling
behavior of the
chiral expansion. We discuss this result in terms of the $1/N_c$ expansion.
We also
consider the implications for determination of the strange quark
contribution to the
nucleon spin from polarized deep inelastic scattering data.

\medskip
PACS Indices: 12.39.Fe, 13.30.Ce, 14.20.Jn

\end{abstract}
\vspace{0.3cm}
]

\pagenumbering{arabic}


\section{Introduction}
\label{sec1}

The chiral expansion of the octet baryon axial current $J_{\mu 5}^A$ has
been a topic of ongoing theoretical
interest for some time. At ${\cal O}(p^0)$, this current is parameterized
by the well-known SU(3) reduced
matrix elements $D$ and $F$. The leading chiral corrections, which arise at
${\cal O}(p^2)$ contain chiral
logarithms, which were first computed in Ref. \cite{wise,john}. Subsequently,
the
wave function  renormalization correction was added in the framework of
heavy baryon
chiral perturbation theory (HBCPT) \cite{j1,j2,ijmpe}, which provides for a
consistent power
counting. While these corrections are large when only octet baryon
intermediate states are kept\cite{j1},
inclusion of decuplet contributions produces sizeable cancellations,
leading to a significantly
smaller ${\cal O}(p^2)$ effect\cite{j2}. The origin of these cancellations
may be
explained by considering
the large $N_c$ expansion \cite{thooft}, as noted in the work of Refs.
\cite{gervas,dash93,dash94,dai95,dash95}. In terms of this counting,
the ${\cal O}(p^0)$ contribution are of order $N_c$, while the ${\cal
O}(p^2)$ loop corrections are
nominally ${\cal O}(N_c^2)$. As shown in Refs.
\cite{dash93,dash94,dai95,dash95,flor00} ,
however, a spin-flavor symmetry arises at this
order whose effect is to render the ${\cal O}(p^2)$ loop effects of
relative order $N_c^0$. Thus, inclusion
of decuplet contributions is crucial to maintining the correct $N_c$
counting as well as the convergence
properties of the chiral expansion through ${\cal O}(p^2)$.

In a recent paper \cite{zhu}, we have calculated the ${\cal O}(p^3)$
corrections to $J_{\mu 5}^A$
arising from octet baryon intermediate states. These corrections are
entirely of recoil order, scaling
as inverse powers of the baryon mass. In that study, we employed baryon
chiral perturbation theory with infrared regularization \cite{becher},
which effectively resums an infinite tower of recoil corrections. Although
this resummation is necessary to
maintain the analytic properties of the currents for momenta near physical
thresholds, we found that
for $q^2=0$ the sum is dominated by the leading $1/M$ correction which
can be obtained directly in
HBCPT. We also found that the ${\cal O}(p^3)$ corrections were large,
exacerbating the poor convergence
obtained through ${\cal O}(p^2)$ in octet-only calculations. We left open
the question as to the impact
of including decuplet intermediate states, speculating that
large-$N_c$ symmetries can generate cancellations at this order as well.

In the present note, we report on an explicit calculation of the ${\cal
O}(p^3)$
corrections which includes
contributions from the decuplet. We find that -- even under the symmetry
constraints imposed by the
large-$N_c$ expansion -- these corrections are both substantial and devoid
of the cancellations arising at
${\cal O}(p^2)$. In several channels, the ${\cal O}(p^3)$ corrections can
be as large as the ${\cal O}(p^0)$
term, in contrast to the naively expected power suppression by
$(m_K/\Lambda_\chi)^2\times(m_K/M)\sim 1/8$. We also
show that the reduced order in $N_c$ arising from the ${\cal O}(p^2)$
spin-flavor algebra is, in retrospect,
what one might expect from the $N_c$ behavior of the relevant counterterms.
In contrast, the ${\cal O}(p^3)$
loop corrections are finite and entirely non-analytic (in quark mass), so
there exists no
counterterm at this order whose $N_c$ behavior would imply a corresponding
order in $N_c$ for the ${\cal
O}(p^3)$ loop corrections. While this observation does not by itself
explain the apparent breakdown of the
chiral expansion for $J_{\mu 5}^A$ at ${\cal O}(p^3)$, it does suggest that
inclusion of decuplet
intermediate states is not generally sufficient to maintain the proper
scaling behavior of the expansion. As a
practical corollary, we also note that the use of SU(3) chiral perturbation
theory to extract $\Delta s$  --
the strange quark contribution to the nucleon spin -- from polarized deep
inelastic scattering data is
subject to uncontrolled approximations and, therefore, untrustworthy.


\section{Axial Currents}
\label{sec3}

In writing down the octet axial currents, it is convenient to start
with the relativistic meson-baryon Lagrangian. At the lowest order, one has
\begin{eqnarray} \nonumber
&{\cal L}_0=i\hbox{ Tr}\left({\bar B} (\gamma^\mu D_\mu -m_N)B \right)+ D
\hbox{ Tr}\left({\bar B}\gamma^\mu\gamma_5 \{A_\mu, B\}\right) \\ \nonumber
&+F\hbox{ Tr}\left({\bar B}\gamma^\mu\gamma_5 [A_\mu, B]\right)
+i{\bar T}^\mu \gamma^\nu D_\nu T_\mu -m_T{\bar T}^\mu  T_\mu \\ \nonumber
&+{\cal C} [{\bar T}^\mu A_\mu B +\bar B A_\mu T^\mu]
+{\cal H}{\bar T}^\mu \gamma^\nu \gamma_5 A^\nu  T_\mu\\
& +{F_\pi^2\over4}\hbox{Tr}\left((D^\mu\Sigma)^{\dag}D_\mu\Sigma\right)
+a \hbox{ Tr} M (\Sigma +\Sigma^\dag ),
\label{lag0}
\end{eqnarray}
where
\begin{eqnarray} \nonumber
&D_\mu\ B=\partial_\mu B +[V_\mu,  B], \\  \nonumber
&D_\mu T^\nu_{abc}=\partial_\mu T^\nu_{abc} +(V_\mu)^d_a
T^\nu_{dbc}+(V_\mu)^d_b T^\nu_{adc}+(V_\mu)^d_c T^\nu_{abd}, \\  \nonumber
&V_\mu={1\over 2}(\xi \partial_\mu \xi^\dag +\xi^\dag \partial_\mu \xi ) \\
\nonumber
&A_\mu={i\over 2}(\xi \partial_\mu \xi^\dag -\xi^\dag \partial_\mu \xi ) \\
\nonumber
&\xi =e^{i{\pi\over F_\pi}}, \   \   \  \Sigma=\xi^2=e^{2i{\pi\over
F_\pi}}, \\  \nonumber
&\pi={1\over \sqrt{2}}\left(\begin{array}{lll}
{\pi^0\over \sqrt{2}}+ {\eta\over \sqrt{6}} &\pi^+&K^+\\
\pi^-&-{\pi^0\over \sqrt{2}}+ {\eta\over \sqrt{6}}&K^0\\
K^-&{\bar K}^0&-{2\over \sqrt{6}}\eta
\end{array}\right) \\  \nonumber
&B=\left(\begin{array}{lll}
{\Sigma^0\over \sqrt{2}}+ {\Lambda\over \sqrt{6}} &\Sigma^+&p\\
\Sigma^-&-{\Sigma^0\over \sqrt{2}}+ {\Lambda\over \sqrt{6}}&n\\
\Xi^-& \Xi^0&-{2\over \sqrt{6}}\Lambda
\end{array}\right) \\ \nonumber
&M=\left(\begin{array}{lll}m_u&0&0\\
0&m_d&0\\0&0&m_s
\end{array}\right)
\end{eqnarray}

One may obtain vector and axial vector current operators from ${\cal L}_0$
by including
vector and axial vector sources in the covariant derivatives. The leading
[${\cal O}(p^0)$] operator
contains only baryon fields and the SU(3) reduced matrix elements $D$ and
$F$. Axial currents
involving both baryons and
mesons first appear at ${\cal O}(p)$. Additional purely baryonic axial
currents appear at ${\cal O}(p^2)$ \cite{zhu}.
They arise from the SU(3) symmetry breaking (SB) Lagrangian
\begin{eqnarray}\label{count} \nonumber
&{\cal L}_1= {m_K^2\over \Lambda_\chi^2}\{
d_1 \hbox{ Tr}\left({\bar B}\gamma^\mu\gamma_5 \{A_\mu, \chi_+\}B\right) \\
\nonumber
&+d_2\hbox{ Tr}\left({\bar B}\gamma^\mu\gamma_5  A_\mu B
\chi_+\right) \\ \nonumber
& +d_3 \hbox{ Tr}\left({\bar B}\gamma^\mu\gamma_5 \chi_+ B A_\mu\right)  \\
&+d_4 \hbox{ Tr}\left({\bar B}\gamma^\mu\gamma_5 B\{A_\mu, \chi_+\}\right)
\},
\label{lag1}
\end{eqnarray}
where
\begin{eqnarray}\nonumber
&\chi_+={1\over 2}(\xi^+ \chi\xi^++\xi\chi^+\xi) \\ \nonumber
&\chi=\left(\begin{array}{lll}0&0&0\\
0&0&0\\0&0&1
\end{array}\right)
\end{eqnarray}

Using ${\cal L}_{0,1}$ one obtains the axial current:
\begin{eqnarray}
J^A_\mu =&
{1\over 2} D\hbox{ Tr}\left({\bar B}\gamma_\mu\gamma_5 \{\xi T^A\xi^\dag
+\xi^\dag
T^A\xi, B\}\right) \nonumber \\
&+{1\over 2} F\hbox{ Tr}\left({\bar B}\gamma_\mu\gamma_5 [\xi T^A\xi^\dag
+\xi^\dag
T^A\xi, B]\right)\nonumber \\
&+{1\over 2}d_1 {m_K^2\over \Lambda_\chi^2}
 \hbox{ Tr}\left({\bar B}\gamma^\mu\gamma_5 \{\xi T^A\xi^\dag +\xi^\dag
T^A\xi, \chi_+\}B\right) \nonumber \\
&+{1\over 2}d_2 {m_K^2\over \Lambda_\chi^2}
\hbox{ Tr}\left({\bar B}\gamma^\mu\gamma_5 (\xi T^A\xi^\dag +\xi^\dag
T^A\xi ) B
\chi_+\right)\nonumber \\
&+{1\over 2}d_3 {m_K^2\over \Lambda_\chi^2}
\hbox{ Tr}\left({\bar B}\gamma^\mu\gamma_5 \chi_+ B(\xi T^A\xi^\dag
+\xi^\dag T^A\xi
)\right)\nonumber \\
&+{1\over 2}d_4 {m_K^2\over \Lambda_\chi^2}
 \hbox{ Tr}\left({\bar B}\gamma^\mu\gamma_5 B\{\xi T^A\xi^\dag +\xi^\dag
T^A\xi, \chi_+\}\right) \nonumber \\
&+{1\over 2} \bar T^\nu \gamma_\mu \left(\xi T^A\xi^\dag
-\xi^\dag T^A\xi\right) T_\nu \nonumber\\
&+{1\over 2} {\cal H} \ \bar T^\nu \gamma_\mu \gamma_5 \left(\xi T^A\xi^\dag
+\xi^\dag T^A\xi\right) T_\nu\nonumber\\
&+{1\over 2} {\cal C} \ \bar T_\mu  \left(\xi T^A\xi^\dag
+\xi^\dag T^A\xi\right) B\nonumber\\
&+{1\over 2} {\cal C} \ \bar B \left(\xi T^A\xi^\dag
+\xi^\dag T^A\xi\right) T_\mu\nonumber\\
 &+{1\over 2}\hbox{ Tr}\left({\bar B}\gamma_\mu [\xi T^A\xi^\dag
-\xi^\dag T^A\xi, B]\right) \nonumber \\
&+{i\over 2}F_\pi^2\hbox{Tr}\ T^A \left((\partial_\mu\Sigma)^{\dag}\Sigma-
\partial_\mu\Sigma \Sigma^+\right) .
\label{current}
\end{eqnarray}

The heavy baryon expansion of ${\cal L}_{0,1}$ and $J_\mu^A$ is obtained
by defining the heavy baryon field $H(x) =exp(im_N  {v\cdot x}) {1+{\hat v}\over 2}B(x)$ 
($v_\mu$ is the baryon velocity) and projecting out the postive
energy states as in \cite{j1}. In this case, all baryon mass terms are
removed from the Lagrangian at leading order, leaving only a dependence 
on the octet-decuplet mass splitting, $\delta=m_\Delta -m_N$. At subleading 
orders, there are recoil corrections in the form of $1/m_N$.
In order to consistently include the decuplet we
follow the small scale expansion proposed in \cite{hb}. In this
approach the energy and momenta and the decuplet and octet mass difference
$\delta$ are both treated as small expansion parameters in chiral counting.
Note also that in the heavy baryon
expansion, one makes the replacement $\gamma_\mu\to v_\mu$,
$\gamma_\mu\gamma_5\to 2 S_\mu$ etc, where $S_\mu$ is the spin operator.

Renormalized matrix elements of $J^A_{\mu 5}$ between octet baryon states
up to ${\cal O}(p^3)$
may be written as
\begin{eqnarray}\label{ren} \nonumber
&\langle B_i| J^A_\mu | B_j\rangle =
\{ \alpha_{ij} +{\bar \alpha}_{ij}{m_K^2\over \Lambda_\chi^2} \\ \nonumber
&+\sum\limits_{X=\pi, K, \eta} [\left(
\lambda_{ij}^X I_d^X+\bar\lambda_{ij}^X I_e^X \right)\alpha_{ij}\\ \nonumber
&+\left(\beta_{ij}^X
I_a^X +\bar\beta_{ij}^X I_f^X +{\tilde\beta}_{ij}^X I_g^X\right) \\
&+\gamma_{ij}^X
I_b^X\alpha_{ij} +\theta_{ij}^X I_c^X \alpha_{ij} ]
 \}
{\bar u}_{B_i}\gamma_\mu\gamma_5 u_{B_j}
\end{eqnarray}
where the first term on the right hand side is the lowest order one.
The second term arises from the SB terms in Eq. (\ref{count}).
The third term in Eq. (\ref{ren}) arises from the wave function
renormalization.
The fourth term comes from the vertex correction diagram.
The fifth term is the vertex correction from the tadpole diagram.
The last term in Eq. (\ref{ren}) arises from the ${\cal O}(p)$ one-meson
operators in $J^A_\mu$. Details of the last three terms can be found in
\cite{zhu}.

Terms with $\alpha_{ij}^X, \bar\alpha_{ij}^X, \lambda_{ij}^X$,
$\beta_{ij}^X$,
$\gamma_{ij}^X$ and
$\theta_{ij}^X$  arise from the contribution of octet states only and their
expressions  are given in \cite{zhu}. The remaining terms come from the
insertion of
decuplet  states in the loop. The expressions of $\bar\lambda_{ij}^X,
\bar\beta_{ij}^X, {\tilde \beta}_{ij}^X$ are presented in Tables \ref{tab1},
\ref{tab2}, and \ref{tab3} respectively.

The functions $I_a^X$ etc are defined as\footnote{Here, $\mu \sim 1$ GeV 
denotes the renormalization
scale. In our previous analysis \cite{zhu}, this scale was effectively set
equal to
$m_N$. Moreover, in that work, the variable $\mu$ denoted the ratio
$m_\pi/m_N$ and
not the renormalization scale.}
\begin{eqnarray}\label{func}\nonumber
I_a^X=({m_X\over \Lambda_\chi})^2\ln ({\mu\over m_X})^2
+\pi {m_X^3\over m_N\Lambda_\chi^2} \\ \nonumber
I_b^X=-({m_X\over \Lambda_\chi})^2\ln ({\mu\over m_X})^2 \\ \nonumber
I_c^X={\pi\over 2} {m_X^3\over m_N\Lambda_\chi^2} \\
I_d^X={3\over 4} I_a^X
\end{eqnarray}
For $X=K, \eta$ we have
\begin{eqnarray}\label{ie}\nonumber
&I_e^X=2{{\cal C}^2\over \Lambda_\chi^2}\{
-[(2\delta^2-m_X^2)\ln ({\mu\over m_X})^2  \\ \nonumber
&+4\delta \sqrt{m_X^2-\delta^2}
Arccos {\delta\over m_X}] \\  \nonumber
&+{1\over m_N}[{2\over 3}(m_X^2-4\delta^2) \sqrt{m_X^2-\delta^2}
Arccos {\delta\over m_X}\\
& +\delta (m_X^2-{4\over 3}\delta^2) \ln ({\mu\over m_X})^2 ]\}
\end{eqnarray}
\begin{eqnarray}\label{if}\nonumber
&I_f^X={20\over 27}{{\cal H}{\cal C}^2\over \Lambda_\chi^2}\{
-[(2\delta^2-m_X^2)\ln ({\mu\over m_X})^2 \\ \nonumber
&+4\delta \sqrt{m_X^2-\delta^2}
Arccos {\delta\over m_X}] \\  \nonumber
&+{1\over m_N}[{2\over 3}(m_X^2-4\delta^2) \sqrt{m_X^2-\delta^2}
Arccos {\delta\over m_X} \\
&+\delta (m_X^2-{4\over 3}\delta^2) \ln ({\mu\over m_X})^2 ]\}
\end{eqnarray}
\begin{eqnarray}\label{ig}\nonumber
&I_g^X={2\over 3}{{\cal C}^2\over \Lambda_\chi^2}\{
-[(m_X^2-{2\over 3} \delta^2)\ln ({\mu\over m_X})^2 \\ \nonumber
&+{4\over 3}
{(m_X^2-\delta^2)^{3\over 2}\over \delta}  Arccos {\delta\over m_X}
-{2\over 3}\pi
{m_X^3\over \delta}] \\  \nonumber
&-{1\over m_N}[{4\over 3}(m_X^2-\delta^2)^{3\over 2}
Arccos {\delta\over m_X}  \\
&+\delta (m_X^2-{2\over 3}\delta^2) \ln ({\mu\over m_X})^2 ]\}
\end{eqnarray}
Replacing the combination
\begin{eqnarray}\nonumber
{Arccos {\delta\over m_X}\over \sqrt{m_X^2-\delta^2}}
\end{eqnarray}
in Eqs. (\ref{ie}-\ref{ig}) by
\begin{eqnarray}\nonumber
{1\over \sqrt{\delta^2-m_X^2}}
\ln \left( {\delta+\sqrt{\delta^2-m_X^2}\over m_X}\right)
\end{eqnarray}
we obtain
expressions for $I_{e,f,g}^\pi$.
In this work we explicitly keep pion loop contribution.
If we truncate at order ${\cal O}(p^2)$ and ignore the pion loops and take
$\delta =0$ and
$m_\eta^2={4\over 3}m_K^2$, we reproduce the expressions in \cite{j1,j2}
exactly. Note that we retain only loop corrections having non-analytic 
dependence on quark masses. Analytic terms (e.g, $\propto m_K^2$) have been 
absorbed into the counterterms $d_{1-4}$. 

\section{$N_c$ counting}
\label{sec3a}

As discussed in a beautiful series of papers
\cite{dash93,dash94,dai95,dash95,flor00}, the
baryon axial currents have
an expansion in $1/N_c$ involving SU(6) spin-flavor operators:
\begin{eqnarray}
G^{ia}&=&q^\dag\frac{\sigma^i}{2}{\lambda^a\over 2} q \\
T^a & = & q^\dag\frac{\lambda^a}{2} q \\
J^i & = & q^\dag\frac{\sigma^i}{2} q \ \ \ ,
\end{eqnarray}
where $q$ and $q^\dag$ are SU(6) quark creation and annihilation operators
and
$\lambda^a$ and
$\sigma^i$ are the Gell-Mann and Pauli matrices, respectively. At leading
order in $1/N_c$,
one has
\begin{equation}\label{eq12}
J_{i5}^a \equiv A^{ia} \propto G^{ia}\ \ \ ,
\end{equation}
where the coefficient of proportionality is of order unity and where terms
of relative order $1/N_c$ have
been dropped. The $N_c$ counting rules give $G^{ia}\sim N_c$. Thus, the
${\cal O}(p^0)$
current is ${\cal O}(N_c)$, while loop corrections, which contain three
insertions of $A^{ia}$ divided by
$F_\pi^2\sim N_c$ are nominally of order $N_c^2$. However, the SU(6)
commutator algebra
\begin{equation}
\left[G^{ia},G^{jb}\right]=\frac{i}{4}\delta^{ij}f^{abc}T^c+\frac{i}{6}\delta^{a
b}\epsilon^{ijk}J^k
+\frac{i}{2}d^{abc}\epsilon^{ijk}G^{kc}
\end{equation}
implies that the ${\cal O}(p^2)$ loop corrections, which depend on double
commutators of $A^{ia}$, are
actually of order $N_c^0$, since each commutator reduces the naive counting
by one power of $N_c$.

Because the ${\cal O}(p^2)$ loops are divergent, there must exist
counterterms of the same
order which absorb the infinities. The most general ${\cal O}(p^2)$
operators arising
at this order include those proportional to $d_{1\ldots 4}$ in Eq.
(\ref{count})
\footnote{There exist additional operators proportional to $m_\pi^2$ and
$q^2$ as well. The
finite parts of the former are numerically insignificant while the latter do
not
contribute to the $q^2=0$ currents. Thus, we do not show them explicitly,
though their
presence is required to remove the divergences.}. These operators involve
one insertion of
$A^{ia}$ times $m_P^2/\Lambda_\chi^2$, where $m_P$ is the Goldstone boson
mass. The latter
is ${\cal O}(N_c^0)$ whereas $\Lambda_\chi^2=(4\pi F_\pi)^2$ is ${\cal
O}(N_c)$.
Thus, the ${\cal O}(p^2)$ counterterms are ${\cal O}(N_c^0)$.
Self-consistency of the
theory implies that the ${\cal O}(p^2)$ loop corrections must also be of
${\cal O}(N_c^0)$.
Otherwise, there would exist a
mismatch between the divergent loops and the counterterms which render them
finite in the
large $N_c$ limit. In retrospect, then, one might have anticipated the
existence of a
large-$N_c$ spin-flavor algebra whose affect is to reduce the nominal $N_c$
order of
the loops to match that of the counterterms.

In contrast, there exist no counterterm operators
at ${\cal O}(p^3)$, and the loop
contributions of this order are entirely finite and non-analytic in $m_q$.
Thus, one has no self-consistency requirement at ${\cal O}(p^3)$  involving
counterterms and loops to force a reduction in the nominal $N_c$ order of
the latter.
In particular, the ${\cal O}(p^3)$ wave function renormalization and vertex
corrections involve three insertions of $A^{ia}$ divided by $F_\pi^2 \times
m_N$.
Since $m_N$ is ${\cal O}(N_c)$, these loop effects are nominally order $N_c$.
In the absence of any algebra which reduces this nominal order, one might
expect
them to be numerically significant. As a practical matter, we find that
inclusion
of decuplet intermediate states produces no cancellations indicative of an
algebraic reduction in the nominal $N_c$ order of these graphs.
Similarly, the ${\cal O}(p^3)$ seagull graphs involving the chiral connection
--which have nominal chiral order ${\cal O}(N_c^0)$--receive only octet
contributions, so no cancellation are possible in this case. We also find
these
contributions can be significant. Indeed, as we show below,
the ${\cal O}(p^3)$
contributions are
generally as large or larger than the
${\cal O}(p^2)$ terms, in accordance with naive scaling arguments.


\section{Numerical Analysis}
\label{sec4}

In tables IV-VII we present various fits to the octet axial currents,
showing the
contributions arising at various orders in $p$. For notational simplicity,
we define
the axial couplings $g^A_{ij}$ as
\begin{equation}
\langle B_i| J_\mu^A | B_j\rangle = g^A_{ij}\ {\bar
u}_{B_i}\gamma_\mu\gamma_5 u_{B_j}
\end{equation}
where we have omitted the induced pseudoscalar terms. In general, we have
eight
low-energy constants (LEC's) to be determined: $D$, $F$, $d_{1-4}$, ${\cal
H}$ and ${\cal
C}$. However, there exist experimental data for only six octet matrix
elements \cite{pdg}.
Consequently, we must invoke additional assumptions in order to complete the
analysis.

The constants ${\cal C}$ and ${\cal H}$ can be treated using one of several
approaches.
Drawing entirely on experimental data, the magnitude of ${\cal C}$ can be
determined from
the decay width of the $\Delta$. At leading order, one has $|{\cal C}| =
1.5$ \cite{hb}, which
is consistent with the large $N_c$ prediction \cite{dai95,dash95}. Loop
corrections to
this result arise
at ${\cal O}(p^2)$. Since ${\cal C}$ enters the axial currents at ${\cal
O}(p^2)$, chiral
corrections to the value of ${\cal C}$ as determined from the $\Delta$ decay
width affect
our analysis at ${\cal O}(p^4)$. Unfortunately, the phase of ${\cal C}$
cannot be
determined in this manner, and so one must rely on auxiliary considerations.
For example,
SU(6) symmetry implies ${\cal C}$ and $D$ have the opposite phase. In what
follows, we
make this choice for the phase.

The situation regarding ${\cal H}$ is more problematic. This LEC does not
appear at leading
order in any physical decay amplitude. It does, however, give the strong
$\pi\Delta\Delta$
coupling at leading order \cite{hb}. A determination of this constant is,
therefore,
highly-dependent on model assumptions. In the
large $N_c$ limit, for example, ${\cal H}=-{9\over 5}(D+F)$. Various quark
models yield the
same result \cite{quark,su6,u12}.  On the other hand, a light cone QCD sum
rule
analysis \cite{zhu1} yields $|{\cal H}|=1.35$, which is only of half of
large
$N_c$ or quark model prediction and is approximately the same value as
extracted from
from the isobar production experiments in $\pi^-p\to \pi^+\pi^-n$ near
threshold \cite{mit}. 
This constant has also been extracted from decay widths using HBCPT 
to ${\cal O}(p^2)$ \cite{but}.
Recently, ${\cal H}$ was determined from a fit to
phase shift data in the fourth order chiral perturbation theory analysis
\cite{mei}. The results imply $0.94 \leq {\cal H} \leq 2.65$. While the
magnitude of ${\cal H}$ for this range
is consistent with both the large $N_c$ and QCD sum rule analyses, the phase
differs
from all other approaches. It was emphasized in Ref. \cite{mei}, however,
that ${\cal H}$
enters pion nucleon scattering at third order loop so it can't be pinned down
precisely.
Fortunately, in the case of the axial currents, ${\cal H}$ arises at ${\cal
O}(p^3)$, so
the impact of uncertainty in this constant is not as pronounced as in the
case of
${\cal C}$.

A final possibility for treating ${\cal C}$ and ${\cal H}$ is to follow the
analysis of
Ref. \cite{dash93,dash94,dai95,dash95,flor00} and invoke the SU(6)
relations: ${\cal C}=-2D$, ${\cal H}=-3D$\footnote{We have also used
the relations arising from the inclusion of $1/N_c$ corrections
to Eq. (\ref{eq12}): ${\cal C}=-2D$, ${\cal H}=3D-9F$
in our fit. The fit results turn out to be the same. We thank E. Jenkins
for suggesting this point.}.
Doing so reduces the number of fit parameters to six
\footnote{A further reduction in number of parameters may occur when
the double expansion in $m_q$ and $1/N_c$ of Ref. \cite{jj} is used.}. The
authors of Refs.
\cite{dash93,dash94,dai95,dash95,flor00} found that use of SU(6) relations
among the
LEC's minimizes the size of the ${\cal O}(p^2)$ loop corrections, in
accordance with the
cancellations expected from large $N_c$ arguments. It is not possible to
apply similar
relations to $d_{1-4}$, however, since they parameterize explicit
symmetry-breaking terms
in the Lagrangian.

In Table IV, we give a fit through ${\cal O}(p^2)$ using the
experimentally-determined
magnitude for ${\cal C}$, a phase opposite to that of $D$, and the quark
model value for
${\cal H}$. The remaining six LEC's are determined from the nucleon and
hyperon
semileptonic decay data. Under these conditions, the ${\cal O}(p^2)$
corrections (including
both loop effects and symmetry breaking terms) are generally as large as the
${\cal
O}(p^0)$ contributions. However, invoking the SU(6) relations among $D$,
${\cal C}$, and
${\cal H}$ changes this situation considerably, as illustrated in Table V.
In this case,
the relative importance of the
${\cal O}(p^2)$ terms is considerably reduced and the $\chi^2$ improved. In
Table VI, we
show the corresponding fit using the SU(6) relations but setting
$d_{1-4}=0$. The latter
fit corresponds roughly to the analysis of Ref.
\cite{dash93,dash94,dai95,dash95,flor00},
which illustrated the impact of ${\cal O}(p^2)$ loop cancellations in the
symmetry limit.
Generally speaking, inclusion of $d_{1-4}$ improves the quality of the fit
as well as the
scaling behavior of the chiral expansion through
${\cal O}(p^2)$.

In Table VII we give the best fit through ${\cal O}(p^3)$. Here, we have
used the SU(6)
relations for $D$, ${\cal C}$, and ${\cal H}$ in order to produce the
cancellations at
${\cal O}(p^2)$. We observe that the ${\cal O}(p^3)$ contributions are
generally as
large or larger than the ${\cal O}(p^2)$ terms and, in several channels, as
large as the
${\cal O}(p^0)$ terms. This pattern becomes even more pronounced away from
the SU(6)
limit for the LEC's, in which case neither the ${\cal O}(p^2)$ nor the
${\cal O}(p^3)$
terms scale as expected.

The breakdown of the chiral expansion which we observe at at ${\cal O}(p^3)$
reflects a number of factors: the large magnitude of the kaon mass, which
appears in the
numerator of the expressions in Eqs. (\ref{func}); the apparent absence of
cancellations (and
an underlying large $N_c$ spin-flavor algebra) among the recoil order
corrections; and
the appearance of factors of $\pi$ in integrals $I_a^X$ and $I_c^X$ arising
at this
order.

\section{Discussion}
\label{sec5}

It has been known for many years that tree-level SU(3) relations are
remarkably successful
in describing a number of the low-lying properties of hadrons, such as
pseudoscalar masses
and baryon axial currents. Ideally, chiral perturbation theory -- together
with the
large $N_c$ expansion -- should suffice explain why these relations work so
well. With such
an understanding in hand, one would have had considerable confidence in
exploiting these
relations to determine quantities for which one has no direct measurement,
such as the
strange quark contribution to the nucleon spin, $\Delta s$. In the present
study, however,
we observe that the chiral expansion for baryon octet axial currents does
not appear to be
under control. While large
$N_c$ considerations imply that the expansion works reasonably well through
${\cal
O}(p^2)$, it breaks down completely at ${\cal O}(p^3)$\footnote{However, we
observe that at ${\cal O}(p^2)$, there exist some channels for which the large
$N_c$ cancellations are not strong (see, e.g., Table \ref{tabb}).}. 
While a theoretical
justification
for applying SU(3) symmetry to the octet axial currents may
exist\footnote{See, {\em
e.g.}, the regulator scheme proposed in Ref. \cite{dhb}}, we are unable to
provide one at this time.

As a practical consequence of this situation, we consider the determination
of $\Delta s$
from polarized deep inelastic scattering (DIS) data.  As shown in
Ref.\cite{ellis}, one may express $\Delta s$ in terms of
the polarized structure
function integrals
\begin{equation}
\Gamma_{p,n}=\int_0^1\ dx\ g_1^{p,n}(x) \ \ \
\end{equation}
as
\begin{equation}
\label{eq:dels}
\Delta s = \frac{3}{2}[\Gamma_p+\Gamma_n]-\frac{5\sqrt{3}}{6} g^A_8
\end{equation}
where $g^A_8$ is the axial vector coupling associated with the matrix
element
$\langle p| J_\mu^8 |p\rangle$. The combinations of LEC's required for this
matrix element are
\begin{eqnarray} \nonumber
\alpha_{pp}^8={1\over 2\sqrt{3}}(3F-D) \\ \nonumber
\beta_{pp}^{8,\pi}={\sqrt{3}\over 8}(3F-D)(D+F)^2 \\ \nonumber
\beta_{pp}^{8,K}={1\over \sqrt{3}}({2\over 3}D^3-2D^2F) \\ \nonumber
\beta_{pp}^{8,\eta}={1\over 24\sqrt{3}}(3F-D)^3 \\ \nonumber
\bar\beta_{pp}^{8,\pi}={\sqrt{3}\over 2} \\ \nonumber
\bar\beta_{pp}^{8, K, \eta}=0 \\ \nonumber
{\tilde\beta}_{pp}^{8,\pi, \eta}=0 \\ \nonumber
{\tilde\beta}_{pp}^{8,K}={\sqrt{3}\over 2}(D-F) \\ \nonumber
{\bar\alpha}_{pp}^8={1\over \sqrt{3}}({1\over 2}d_2-2d_4) \\ \nonumber
\gamma_{pp}^{8,K}=-{3\over 2},\; \; \gamma_{pp}^{8, \pi, \eta}=0 \\
\nonumber
\theta_{pp}^{8,\pi, K, \eta}=-4\gamma_{pp}^{8,\pi, K, \eta} \nonumber
\end{eqnarray}

The numerical separation of $g^A_8$ through ${\cal O}(p^3)$ is given in
Table VII and
yields
\begin{equation}
\Delta s = 0.14 - [0.12 + 0.25 + 0.10] \ \ \ ,
\end{equation}
where the numbers in square brackets correspond, respectively, to the order
$p^0$, $p^2$,
and $p^3$ contributions to $g^A_8$.  Since the chiral expansion is not
converging
for $\Delta s$, we do
not quote a total for this quantity nor can we estimate a theoretical
uncertainty. In
contrast, extractions of $\Delta s$ from semi-inclusive
measurements performed by the Hermes collaboration \cite{hermes} or from
elastic
neutrino-nucleon scattering \cite{garvey,mu90} are not plagued by large
SU(3)-breaking
uncertainties,
making them in principle more reliable probes of the flavor content of the
nucleon spin.

This work was supported in part under U.S. Department of
Energy contract \#DE-FGO2-00ER41146,
the National Science Foundation under award PHY00-71856,
and a National Science Foundation Young Investigator Award.
We thank E. Jenkins, B. R. Holstein, and J. L. Goity for useful discussions.


\begin{table}
\begin{center}~
\begin{tabular}{|c|c|c|c|}\hline
 & $\pi$ loop & kaon loop&$\eta$ loop
\\ \hline

$\bar\lambda_{pn}$ & $1$  & ${1\over 4}$ &$0$  \\ \hline

$\bar\lambda_{\Lambda\Sigma^-}$ & ${11\over 24}$ & ${2\over 3}$ &${1\over
8}$\\ \hline

$\bar\lambda_{\Xi^0\Xi^-}$ & ${1\over 4}$ & ${3\over 4} $&${1\over 4}$\\
\hline

$\bar\lambda_{p\Lambda}$ & ${7\over 8}$ & ${3\over 8} $&$0$\\ \hline

$\bar\lambda_{\Lambda\Xi^-}$ & ${1\over 2} $ & ${5\over 8} $ &${1\over 8}$
\\ \hline

$\bar\lambda_{n\Sigma^-}$  & ${7\over 12}$ & ${13\over 24} $ &${1\over 8}$
\\ \hline

$\bar\lambda_{\Sigma^0\Xi^-}$  & ${5\over 24}$ & ${19\over 24}$ &${1\over
4}$ \\
\hline

$\bar\lambda_{pp}$  & $1$ & $ {1\over 4}$ & $0$\\ \hline

$\bar\lambda_{\Lambda\Lambda}$  & ${3\over 4}$ & ${1\over 2} $ &$0$\\ \hline

$\bar\lambda_{\Sigma\Sigma}$  & ${1\over 6}$ & ${5\over 6}  $ &${1\over
4}$\\ \hline

$\bar\lambda_{\Xi\Xi}$  & ${1\over 4}$ & ${3\over 4} $ &${1\over 4}$\\
\hline
\end{tabular}
\end{center}
\caption{\label{tab1} The coefficients $\bar\lambda^X_{ij}$ for the wave
function
renormalization due to the decuplet intermediate states.}
\end{table}

\begin{table}
\begin{center}~
\begin{tabular}{|c|c|c|c|}\hline
 & $\pi$ loop & kaon loop&$\eta$ loop \\ \hline
$\bar\beta_{pn}$ & ${5\over 6}$ & ${1\over 6}$ &$0$ \\ \hline

$\bar\beta_{\Lambda\Sigma^-}$ & ${1\over 2\sqrt{6}}$ & ${1\over 4\sqrt{6}}$
&$0$ \\
\hline

$\bar\beta_{\Xi^0\Xi^-}$ & ${1\over 24}$ & $-{1\over 6}$ &$-{1\over 8}$ \\
\hline

$\bar\beta_{p\Lambda}$ & $-{\sqrt{6}\over 4}$ & $-{\sqrt{6}\over 8}$ &$0$
\\ \hline

$\bar\beta_{\Lambda\Xi^-}$ & ${\sqrt{6}\over 8}$ & ${\sqrt{6}\over 8}$ &$0$
\\ \hline

$\bar\beta_{n\Sigma^-}$  & $-{1\over 6}$ & $-{1\over 12}$ &$0$ \\ \hline

$\bar\beta_{\Sigma^0\Xi^-}$  & ${1\over 6\sqrt{2}}$ & ${7\over 12\sqrt{2}}$
&${1\over
4\sqrt{2}}$ \\ \hline
\end{tabular}
\end{center}
\caption{\label{tab2} The coefficients $\bar\beta^X_{ij}$ for the vertex
corrections.}
\end{table}

\begin{table}
\begin{center}~
\begin{tabular}{|c|c|c|c|}\hline
 & $\pi$ loop & kaon loop&$\eta$ loop \\ \hline
${\tilde\beta}_{pn}$ & ${8\over 3}(D+F)$ & ${F+3D\over 3}$ &$0$ \\ \hline

${\tilde\beta}_{\Lambda\Sigma^-}$ & ${2\over \sqrt{6}}F$ & ${4\over
\sqrt{6}}(F+{2\over
3}D)$ &${1\over \sqrt{6}}D$
\\
\hline

${\tilde\beta}_{\Xi^0\Xi^-}$ & $-{D-F\over 3}$ & ${5F+D\over 3}$
&${3F+D\over 3}$ \\ \hline

${\tilde\beta}_{p\Lambda}$ & $-{1\over 2\sqrt{6}}(3F+11D)$ & $-{3\over
2\sqrt{6}}(F+D)$
&$0$ \\ \hline

${\tilde\beta}_{\Lambda\Xi^-}$ & $-{1\over 2\sqrt{6}}(3F-D)$ & ${3\over
2\sqrt{6}}(D-F)$
&${1\over \sqrt{6}}D$ \\ \hline

${\tilde\beta}_{n\Sigma^-}$  & ${1\over 3}(D+5F)$ & ${1\over 6}(5F+D)$
&${1\over 6}(3F-D)$
\\ \hline

${\tilde\beta}_{\Sigma^0\Xi^-}$  & ${\sqrt{2}\over 6}(2D+F)$ &
${\sqrt{2}\over
12}(15D+13F)$ &${\sqrt{2}\over 4}(D+F)$
\\ \hline
\end{tabular}
\end{center}
\caption{\label{tab3} The coefficients ${\tilde\beta}^X_{ij}$ for the
vertex corrections.}
\end{table}

\begin{table}
\begin{center}
\begin{tabular}{|c|c|c|c|}\hline
 & Full fit results & Tree level only& ${\cal O}(p^2) $ only
\\
\hline
$g^A_{pn}$ &                $1.28 $  & $0.18 $ & $1.10 $ \\ \hline

$g^A_{\Lambda\Sigma^-}$ &   $0.59 $  & $0.51 $ & $0.08 $ \\ \hline

$g^A_{p\Lambda}$ &          $-0.83 $  & $0.29 $ & $-1.12 $ \\ \hline

$g^A_{\Lambda\Xi^-}$ &      $0.29 $  & $-0.81 $ & $1.10 $ \\ \hline

$g^A_{n\Sigma^-}$  &        $0.32 $  & $1.08 $ & $-0.76 $ \\ \hline

$g^A_{\Sigma^0\Xi^-}$  &    $0.97 $  & $0.13 $ & $0.84 $ \\ \hline

$g^A_{\Xi^0\Xi^-}$$^{\dag}$& $-0.02 $  & $1.08 $ & $-1.10 $ \\ \hline

$g^A_8$$^{\dag}$&           $0.32 $  & $-0.57 $ & $0.89 $ \\ \hline
\end{tabular}
\end{center}
\caption{\label{taba} The separation of fit
results into pure ${\cal O}(p^0)$ and ${\cal O}(p^2)$
pieces where we have used $\delta=0.3$ GeV, $m_N\to \infty$,
${\cal C}=-1.5$, ${\cal H}=-2.25$ as inputs. The fit yields $D=0.63,
F=-0.45, d_1=0.79, d_2=1.87,
d_3=1.43, d_4=-1.13$ with $\chi^2=0.15$.
}
\end{table}

\begin{table}
\begin{center}
\begin{tabular}{|c|c|c|c|}\hline
& Full fit results & Tree level only& ${\cal O}(p^2) $ only
\\
\hline
$g^A_{pn}$ &                $1.26 $  & $0.77 $ & $0.49 $ \\ \hline

$g^A_{\Lambda\Sigma^-}$ &   $0.62 $  & $0.38 $ & $0.24 $ \\ \hline

$g^A_{p\Lambda}$ &          $-0.89 $  & $-0.57 $ & $-0.32 $ \\ \hline

$g^A_{\Lambda\Xi^-}$ &      $0.32 $  & $0.19 $ & $0.13 $ \\ \hline

$g^A_{n\Sigma^-}$  &        $0.34 $  & $0.15 $ & $0.19 $ \\ \hline

$g^A_{\Sigma^0\Xi^-}$  &    $0.92 $  & $0.54 $ & $0.38 $ \\ \hline

$g^A_{\Xi^0\Xi^-}$$^{\dag}$& $0.15 $  & $0.15 $ & $ 0$ \\ \hline

$g^A_8$$^{\dag}$&           $0.17 $  & $0.13 $ & $0.04 $ \\ \hline
\end{tabular}
\end{center}
\caption{\label{tabb} The separation of fit
results into pure ${\cal O}(p^0)$ and ${\cal O}(p^2)$
pieces where we have used $\delta=0.3$ GeV, $m_N\to \infty$,
${\cal C}=-2D$, ${\cal H}=-3D$ as inputs. The fit yields $D=0.46,
F=0.31, d_1=-0.80, d_2=0.93,
d_3=-0.63, d_4=0.78$ with $\chi^2=0.002$.
}
\end{table}

\begin{table}
\begin{center}
\begin{tabular}{|c|c|c|c|}\hline
 & Full fit results & Tree level only& ${\cal O}(p^2) $ only
\\
\hline
$g^A_{pn}$ &                $1.10 $  & $0.76 $ & $0.34 $ \\ \hline

$g^A_{\Lambda\Sigma^-}$ &   $0.66 $  & $0.42 $ & $0.24 $ \\ \hline

$g^A_{p\Lambda}$ &          $-0.88 $  & $-0.51 $ & $-0.37$ \\ \hline

$g^A_{\Lambda\Xi^-}$ &      $0.31 $  & $0.10 $ & $0.21 $ \\ \hline

$g^A_{n\Sigma^-}$  &        $0.29 $  & $0.26 $ & $0.03 $ \\ \hline

$g^A_{\Sigma^0\Xi^-}$  &    $1.05 $  & $0.54 $ & $0.51 $ \\ \hline

$g^A_{\Xi^0\Xi^-}$$^{\dag}$& $0.35 $  & $0.26 $ & $0.09 $ \\ \hline

$g^A_8$$^{\dag}$&            $0.26 $  & $0.07 $ & $0.19 $ \\ \hline
\end{tabular}
\end{center}
\caption{\label{tabc} The separation of fit
results into pure ${\cal O}(p^0)$ and ${\cal O}(p^2)$
pieces where we have used $\delta=0.3$ GeV, $m_N\to \infty$,
$d_{1-4}=0$,
${\cal C}=-2D$, ${\cal H}=-3D$ as inputs. The fit yields $D=0.51,
F=0.25$ with $\chi^2=1.1$.
}
\end{table}

\begin{table}
\begin{center}
\begin{tabular}{|c|c|c|c|c|}\hline
 & Full fit results & Tree level only& ${\cal O}(p^2) $ only& ${\cal
O}(p^3)$ only
\\
\hline
$g^A_{pn}$ &                 $1.26 $  & $0.61$ & $0.41$ & $0.24$\\ \hline

$g^A_{\Lambda\Sigma^-}$ &    $0.58 $  & $0.32$ & $0.14$ & $0.12$\\ \hline

$g^A_{p\Lambda}$ &           $-0.92 $  & $-0.43$ & $-0.11$ & $-0.38$\\
\hline

$g^A_{\Lambda\Xi^-}$ &       $0.26 $  & $0.11$ & $0.05$ & $0.10$\\ \hline

$g^A_{n\Sigma^-}$  &         $0.33 $  & $0.17$ & $0.03$ & $0.13$\\ \hline

$g^A_{\Sigma^0\Xi^-}$  &     $0.87 $  & $0.43$ & $0.05$ & $0.39$\\ \hline

$g^A_{\Xi^0\Xi^-}$$^{\dag}$& $0.22 $  & $0.17$ & $-0.02$ & $0.07$\\ \hline

$g^A_8$$^{\dag}$&            $0.32 $  & $0.08$ & $0.17$ & $0.07$\\ \hline
\end{tabular}
\end{center}
\caption{\label{tabg} The separation of fit
results into pure ${\cal O}(p^0)$, ${\cal O}(p^2)$, and ${\cal O}(p^3)$
pieces where we have used $\delta=0.3$ GeV, $m_N=0.94$ GeV,
${\cal C}=-2D$, ${\cal H}=-3F$ as inputs. The fit yields
$D=0.39, F=0.22, d_1=-1.97, d_2=1.14,
d_3=-0.45, d_4= -0.06$ with $\chi^2=0.12$.
}
\end{table}

\end{document}